\documentclass{emulateapj}
\usepackage{textcomp}
\usepackage{mathtext}
\usepackage{amssymb}
\usepackage{rotating}

\newcommand{\fermi}{\textit{Fermi}}
\newcommand{\gr}{$\gamma$-ray}
\newcommand{\psr}{PSR J0614$-$3329}
\newcommand{\jcap}{JCAP}

\begin{document}

\title{\textit{Fermi} Study of $\gamma$-Ray Millisecond Pulsars: the Spectral Shape and Pulsed 25--200 GeV Emission from J0614$-$3329}

\author{Yi Xing and Zhongxiang Wang}

\affil{Key Laboratory for Research in Galaxies and Cosmology,
Shanghai Astronomical Observatory,\\ 
Chinese Academy of Sciences, 80 Nandan Road, Shanghai 200030, China}

\begin{abstract}
We report our analysis of the \textit{Fermi} Large Area Telescope data
for 39 millisecond pulsars (MSPs) listed in the second $\gamma$-ray
pulsar catalog. Spectra of the pulsars are obtained. We fit
the spectra with a function of a power law with exponential cutoff, and
find the best-fit parameters of photon index $\Gamma = 1.54^{+0.10}_{-0.11}$ 
and cutoff energy $E_{c} = 3.70^{+0.95}_{-0.70}$ GeV. This spectral shape,
which includes the intrinsic differences in the spectra of the MSPs,
can be used for finding candidate MSPs and unidentified types of sources 
detected by \textit{Fermi} at high Galactic latitudes. In one of 
the MSPs PSR J0614$-$3329, we find 
significant pulsed emission upto 200 GeV. The result has thus added this MSP
to the group of the Crab and Vela pulsars that have been detected with
$>$50 GeV pulsed emission. Comparing the $\gamma$-ray spectrum of 
PSR~J0614$-$3329 with those of the Crab and Vela pulsars, we discuss possible
emission mechanisms for the very high-energy component.

\end{abstract}


\section{Introduction}

Since the launch of \textit{Fermi Gamma-Ray Space Telescope (Fermi)} in year
2008, the Large Area Telescope (LAT) onboard it 
has been scanning the whole sky with unprecedented sensitivity 
at 0.1--300~GeV energy range. Thus far, more than 3000 \gr\ sources
have been observed at the \gr\ energy range \citep{3fgl15}, and we 
are able to study bright sources among them in great detail. 
From \fermi\ LAT observations, we have
learned that pulsars are the prominent \gr\ sources in our Galaxy.
More than 200 pulsars have been found with \gr\ emission, half of which
are millisecond pulsars (MSPs; \citealt{2fpsr13}).\footnote{https://confluence.slac.stanford.edu/display/GLAMCOG/Public+List+of+LAT-Detected+Gamma-Ray+Pulsars}
Emission from pulsars at the \fermi\ LAT energy range generally can be described
by a power law with exponential cutoff, where the cutoff energy
is in a range of 1--7~GeV \citep{2fpsr13}. This spectral feature, along
with that of stable emission, can be used for finding good pulsar candidates
among the unidentified \gr\ sources found by \fermi.

MSPs are $\sim 10^9$ yr old neutron stars, having evolved from low-mass
X-ray binaries by gaining sufficient angular momentum from 
accretion \citep{alp+82,rs82}. Because of their old ages,
the \gr\ MSPs appear to be isotropically distributed in the sky \citep{2fpsr13}.
The distribution makes them mixed with the extragalactic \gr\ sources, 
which include Active Galactic Nuclei (AGN; the major class of \gr\
sources in the sky), several other types
of galaxies \citep{3fgl15}, and even possibly unidentified types of sources
(e.g., \citealt{bhl15}). With the release of \fermi\ LAT Pass 8 database
in year 2015, the detection sensitivity has been improved significantly, 
particularly at the low and high end of the LAT energy range. More faint 
sources, in addition to $\sim 3000$ sources reported in the LAT third source
catalog \citet{3fgl15}, appear in the data analysis. For the purpose of
finding candidate MSPs (e.g., \citealt{dai+16}), a fine definition for
the spectral shape of \gr\ MSPs is needed. 

We therefore have conducted analysis of the LAT data for 39 \gr\ MSPs 
reported in the second LAT catalog of \gr\ pulsars (hereafter 2PC). We have 
extracted their spectra in a uniform way 
by using the latest Pass 8 database, and obtained the general
spectral shape from their spectra. In addition, our analysis has revisited 
the $>$10\,GeV emission found in three MSPs by \citet{ack10gev+13}, 
and in \psr, we have found significant upto 200~GeV emission. In this paper, 
we report these results.

\section{\textit{Fermi} LAT Data}

LAT is a $\gamma$-ray imaging instrument onboard \fermi\ 
that scans the whole sky every three hours and can continuously 
conduct long-term \gr\ observations of thousands of GeV 
sources \citep{atw+09}. 
In this analysis, we selected 39 of 40 MSPs listed in 2PC \citep{2fpsr13} as 
our targets (see Table~\ref{tab:likelihood}), while PSR~J1939+2134 was 
not included because of the low detection significance for 
it ($\simeq 3\sigma$). 
The data we used for each target are the 0.1--300 GeV LAT events 
in the \textit{Fermi} Pass 8 database inside 
a $\mathrm{20^{o}\times20^{o}}$ region centered at a target's position.
To fully study the very high-energy emission from \psr, for
the detailed data analysis for this MSP, the high-energy
end was extended to 500~GeV.  The time period of the LAT data is 
from 2008-08-04 15:43:36 (UTC) to 2016-01-28 00:08:16 (UTC). Following 
the recommendations of the LAT 
team\footnote{\footnotesize http://fermi.gsfc.nasa.gov/ssc/data/analysis/scitools/}, 
we included those events with zenith angles less than 90 degrees, which 
prevents the Earth's limb contamination, and excluded the events with 
quality flags of `bad'.
In our following analysis, the background Galactic and 
extragalactic diffuse emission of 
the spectral model gll\_iem\_v06.fits and the 
file iso\_P8R2\_SOURCE\_V6\_v06.txt, respectively, were used. 
The normalizations of 
the diffuse components in the analysis were always set as free parameters.
\begin{figure}
\centering
\epsscale{1.0}
\plotone{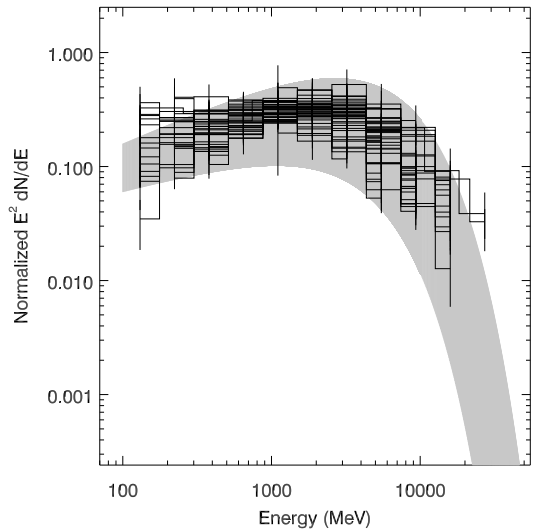}
\caption{Normalized spectra of 39 MSPs. The uncertainties include
the statistic and systematic ones (with the latter due to the Galactic diffuse 
emission model used). The grey area indicates the 3$\sigma$ region of 
the best-fit spectral model.}
\label{fig:spectra}
\end{figure}

\section{Data Analysis and Results for 39 MSPs} 
\label{sec:ana}

\subsection{Likelihood Analysis}
\label{subsec:la}

For each of the MSP targets, we included all sources within 20 degrees 
centered at their positions to make the source models based on 
the \textit{Fermi} LAT 4-yr catalog \citep{3fgl15}. The spectral forms 
of these sources are provided in the catalog. Spectral parameters of 
the sources within 5 degrees from each target were set as free parameters, 
and the other parameters were fixed at their catalog values. 
The catalog spectral models for 33 MSPs are an exponentially cutoff 
power law, $dN/dE = N_{0}E^{-\Gamma}\exp(-E/E_{c})$, while for the 
other six MSPs (J0610$-$2100, J1446$-$4701, J1747$-$4036, J1125$-$5825, 
J1741$+$1351, and J1823$-$3021A) are a simple power law,
$dN/dE = N_{0}E^{-\Gamma}$. 

Using the LAT science tools software package {\tt v10r0p5}, we performed 
standard binned likelihood analysis to the LAT data of the MSP targets in 
the $>$0.1 GeV band. For PSRs J1658$-$5324 and J1858$-$2216, the analysis 
could not converge, which might be because of the relatively large 
uncertainties of the instrument response function of the LAT in the low 
energy range. We thus used $>$0.2 GeV data instead for the two sources. 
The spectral results as well as the Test Statistic (TS) values 
are given in Table~\ref{tab:likelihood} for each source. 
The TS value at a given position is calculated from TS$= -2\log(L_{0}/L_{1})$, 
where $L_{0}$ and $L_{1}$ are the maximum likelihood values for a model 
without and with an additional source respectively. It is a measurement of 
the fit improvement for including the source, and is approximately 
the square of the detection significance of the source \citep{1fgl}.

For the 6 MSPs with a power-law spectral model in the catalog, 
we repeated the analysis with an exponentially cutoff power law.
The significance of a spectral cutoff was estimated from
$\sqrt{-2\log(L_{pl}/L_{exp})}$, where
$L_{exp}$ and $L_{pl}$ are the maximum likelihood values when  
a target's emission was modeled with a power law with and without the cutoff 
respectively \citep{2fpsr13}. 
We found that for the 6 pulsars, the spectral cutoff was detected 
with $>$3$\sigma$ significance.  Therefore in Table~\ref{tab:likelihood},
we only provide the exponentially cutoff power-law results for them.

\subsection{Spectral Analysis}
\label{subsec:sa}

We extracted the $\gamma$-ray spectra of the MSP targets by performing 
maximum likelihood analysis to the LAT data in 15 evenly 
divided energy bands in logarithm from 0.1--300 GeV. 
In the extraction, the spectral normalizations of the sources within 5 
degrees from each target were set as free parameters, while all the other 
parameters of the sources were fixed at the values obtained from the above 
maximum likelihood analysis. The targets were considered as point sources 
having power-law emission with $\Gamma$ fixed at 2.0. The fluxes obtained in 
this way are less dependent on the overall spectral model assumed for a
source, providing a good description for the \gr\ emission of the source. 
We kept only flux data points when TS greater 
than 9 (i.e., $>$3$\sigma$ significance). A total of 304 data points were 
obtained for the 39 targets. The flux values for each target are
provided in Table~\ref{tab:point}.
We also estimated the systematic uncertainties  
caused by the Galactic diffuse emission model used. 
The uncertainty in each energy band was obtained by repeating the 
likelihood analysis with the 
normalization of the diffuse component artificially fixed to the values
$\pm$6\% deviating from the best-fit value
(see, e.g., \citealt{2fpsr13}). 
The uncertainties given in Table~\ref{tab:point} have included 
the systematic uncertainties. We checked the spectrum and best-fit model
for each target. The spectra are well fitted by the spectral models 
obtained from the likelihood analysis.

\subsection{Spectral Shape Determination}

In order to obtain a spectral shape that generally defines emission
from MSPs, we first normalized the fluxes of each MSP target with 
its 0.1--300 GeV energy flux ($F_{100}$ in Table~\ref{tab:likelihood}). 
The normalized spectra of the 39 MSPs are shown in 
Figure~\ref{fig:spectra}. We then fit these data points with a normalized 
exponentially cutoff power law, i.e., $N_{0}$ is obtained from $\Gamma$ 
and $E_{c}$ by requiring the total flux to be 1.
The best-fit values we obtained were $\Gamma$ = 1.5 and E$_{c}$ = 3.8 GeV, but
with a minimum $\chi^{2}$ value of 2198 for 302 degrees of freedom. 
The large $\chi^2$ reflects the intrinsic spectral differences of the MSPs.

We thus used a systematic uncertainty parameter to represent the intrinsic 
differences.  The parameter was added to the uncertainties
of the data points in quadrature. We found that when this parameter was set
to be 0.05, the minimum reduced $\chi^2$ was approximately equal to 1.
As a result, $\Gamma = 1.54^{+0.10}_{-0.11}$ and 
$E_{c} = 3.70^{+0.95}_{-0.70}$ GeV were obtained, where the uncertainties
are at a 3$\sigma$ confidence level.
This 3$\sigma$ spectral region is shown as the grey area in 
Figure~\ref{fig:spectra}.

\section{Data Analysis and Results for \psr} 
\label{sec:anas}

In our analysis, we naturally revisited the high-energy $>$10 GeV emission
from three MSPs found by \citet{ack10gev+13} in 2PC. In \psr, we found
a significant high-energy component and thus conducted detailed analysis
of the data for this pulsar.
\begin{figure}
\centering
\epsscale{1.0}
\plotone{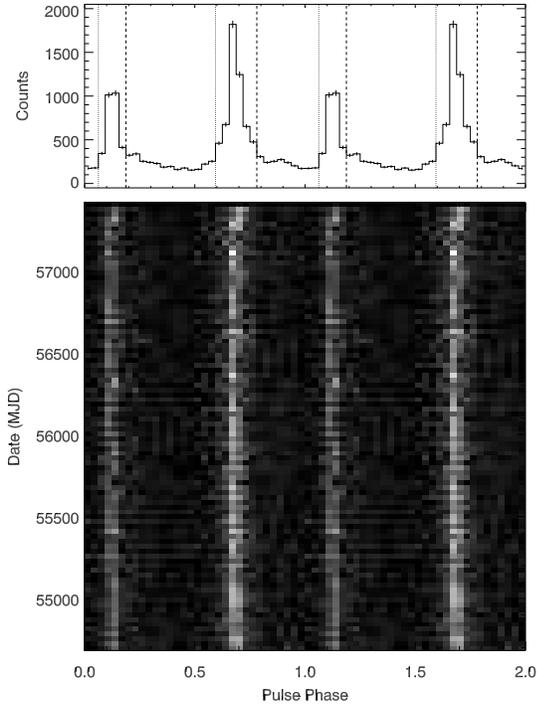}
\caption{Folded pulse profile and two-dimensional phaseogram in 32 phase
bins obtained for \psr. The grayscale represents the number of
photons in each bin. The dotted and dashed lines mark the  
phase ranges defined for the on-pulse phase intervals.
Two cycles are displayed for clarity.}
\label{fig:profile}
\end{figure}

\subsection{Timing Analysis}
\label{subsec:ta}

We performed timing analysis to the 0.1--500 GeV LAT data of 
the \psr\ region to update the \gr\ ephemeris given in \citet{2fpsr13}. 
An aperture radius of 1\fdg0 was used. Pulse phases for photons
before MJD 55797 (the end time of the known ephemeris) were assigned 
according to the known ephemeris using the \fermi\ plugin 
of TEMPO2 \citep{hem06,ehm06}. An `empirical Fourier' template 
profile was built. Using this template, we generated the times of arrival 
(TOAs) of 40 evenly divided observations of the whole time period. Both the template and TOAs 
were obtained using the maximum likelihood method described
in \citet{ray+11}.

We used TEMPO2 to fit the TOAs. Only the pulse frequency derivative 
$\dot{f}$ was fitted, and the other timing parameters were fixed to 
their known values. We obtained $\dot{f}= -1.7559(1)\times 10^{-15}$\,s$^{-2}$,
consistent with the value given in \citet{2fpsr13} 
within $\sim$2.2$\sigma$ uncertainty. The folded pulse profile 
and two-dimensional phaseogram are shown in Figure~\ref{fig:profile}.
In the following analysis, we selected phase 0.06--0.19 and 0.59--0.78 
as the onpulse phase intervals, and the rest as the offpulse phase intervals.

\subsection{Likelihood Analysis}
\label{subsec:las}

We included all sources within 20 degrees centered at the position 
of \psr\ in the \textit{Fermi} LAT 4-year catalog \citep{3fgl15} to make 
the source model. The spectral forms of these sources are provided in 
the catalog. Spectral parameters of the sources within 5 degrees 
from \psr\ were set as free parameters, and the other parameters were 
fixed at their catalog values. The catalog spectral form of \psr\ is an 
exponentially cutoff power law, 
$dN/dE = N_{0}E^{-\Gamma}\exp[-(E/E_{c})^{b }]$.
The parameter $b$ is a measurement of the exponential cutoff shape, where
a value of 1 or $<$1 indicates a simple exponential cutoff or a 
sub-exponential cutoff, respectively. We also used a simple power 
law in the analysis for comparison. 
\begin{figure}
\centering
\epsscale{1.0}
\plotone{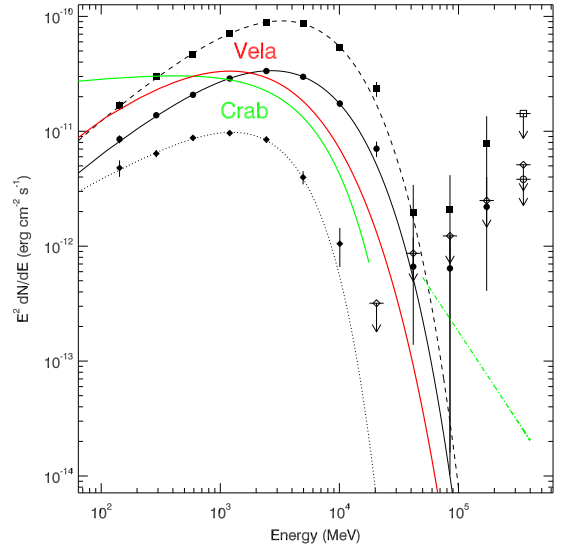}
\caption{\fermi\ \gr\ spectra of \psr\ during the total (dots), 
onpulse (squares), and offpulse (diamonds) phase intervals. 
The solid and dashed curves are the 0.1--500 GeV sub-exponentially 
cutoff power-law fits to emission during the total and onpulse phase intervals, 
respectively. The dotted curve is the 0.1--500 GeV exponentially cutoff 
power-law fit to emission during the offpulse phase intervals. The flux-scaled
model fits to $\gamma$-ray emission of the Crab \citep{2fpsr13,ale+11} 
and Vela \citep{leu+14} pulsars are shown as green and red curves,
respectively, for comparison.}
\label{fig:spectras}
\end{figure}

We performed standard binned likelihood analysis to the LAT data in 
$>$0.1 GeV energy range. We first set $b= 1$. The \gr\ emission 
during the total pulse phase intervals was detected with a TS 
value of 33576, while that during the onpulse and offpulse phase intervals 
were detected with TS values of 37013 and 3766, respectively.
We found that during the total, onpulse, and offpulse phase 
intervals, the emission was better modeled with an exponentially cutoff power 
law. The cutoffs were significantly detected during all the three
phase intervals ($>5\sigma$; estimated from $\sqrt{-2\log(L_{pl}/L_{exp})}$). The resulting power-law fits with simple exponential cutoff
are summarized in Table~\ref{tab:likelihoods}.

We then set $b$ as a free parameter and repeated the binned likelihood analysis 
to the LAT data. We found that during the total and onpulse phase intervals the 
sub-exponential cutoffs were detected with $\sim$4$\sigma$ significance 
(estimated from $\sqrt{-2\log(L_{exp}/L_{subexp})}$, 
where $L_{subexp}$ is the maximum likelihood value for 
the sub-exponentially cutoff power-law model; \citealt{2fpsr13}). 
The resulting sub-exponentially cutoff power-law fits during 
these two phase intervals are given in Table~\ref{tab:likelihoods}. 
During the offpulse phase interval, the sub-exponential cutoff was not 
detected, as the detection significance was approximately zero.
\begin{figure}
\centering
\epsscale{1.0}
\plotone{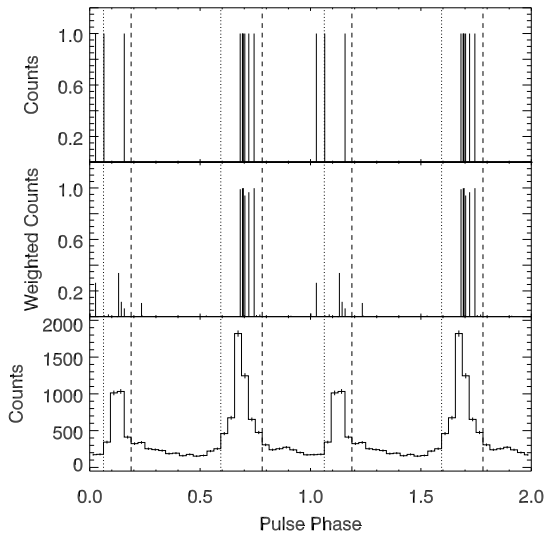}
\caption{{\it Top} panel: $>$25 GeV photons within 0\fdg5 from \psr. 
{\it Middle} panel: weighted $>$25 GeV photons within $\mathrm{2^{o}}$ 
from \psr.
{\it Bottom} panel: 0.1--500 GeV folded pulse profile in 32 phase bins 
obtained for \psr. The dotted and dashed lines mark the phase ranges
for the on-pulse phase intervals. Two cycles are displayed for clarity.}
\label{fig:profile25}
\end{figure}

\subsection{Spectral Analysis}
\label{subsec:sas}

We extracted the $\gamma$-ray spectra of \psr\ during the total, onpulse, 
and offpulse phase intervals, by performing maximum likelihood analysis 
to the LAT data in 12 evenly divided energy bands in logarithm 
from 0.1--500 GeV. In the extraction, the spectral normalizations of the 
sources within 5 degrees from \psr\ were set as free parameters, while 
all the other parameters of the sources were fixed at the values obtained 
from the above maximum likelihood analysis. We kept only spectral flux points 
when TS greater than 4 ($>$2$\sigma$ significance) and derived 95\% flux 
upper limits otherwise. The obtained spectra are shown in 
Figure~\ref{fig:spectras}, and the fluxes and TS values are provided 
in Table~\ref{tab:spectra}. We found that while the offpulse emission was 
detected in an energy range of only $<$15 GeV, 
the onpulse emission from the pulsar was significantly detected 
in a high-energy range, upto approximately 200 GeV 
(see Table~\ref{tab:spectra}).

\subsection{Timing Analysis of $>$25 GeV data}
\label{subsec:tah}

We performed timing analysis to the LAT data of \psr\ to search for 
\gr\ pulsations at the high-energy range, for which we selected 
the minimum energy
as high as possible but also ensured sufficient pulsation detection 
significance. We found the value of 25 GeV used in \citet{ack10gev+13} was 
proper. We first selected \gr\ photons within 
an aperture radius of 0.5 degrees from \psr, approximately corresponding to 
the 95\% contamination angle of the incoming photons from a source. 
A total of ten photons 
were collected. Pulse phases for the photons were assigned using 
the updated ephemeris obtained in Section~\ref{subsec:ta}, 
and an H-test value of 30 was obtained, corresponding to 4.5$\sigma$ 
detection significance \citep{jb10}. 
These photons are shown in the top panel of Figure~\ref{fig:profile25} 
according to their pulse phases, and the 0.1--500 GeV pulse profile is
shown in the bottom panel of Figure~\ref{fig:profile25} for comparison.

We then used a larger aperture radius of 2 degrees to include more 
photons (40 photons were collected), and weighted them by their probability 
of originating from the pulsar (calculated with using \textit{gtsrcprob}). 
Pulse phases for these photons were assigned and a weighted H-test value of 
48 was obtained \citep{jb10,ker11}, corresponding to $\sim$6$\sigma$ detection 
significance, indicating that the \gr\ pulsation from the source was 
significantly detected in $>$25 GeV energy band. We plotted the weighted 
photons in the middle panel of Figure~\ref{fig:profile25} according 
to their pulse phases.

\begin{figure*}
\centering
\epsscale{1.0}
\plotone{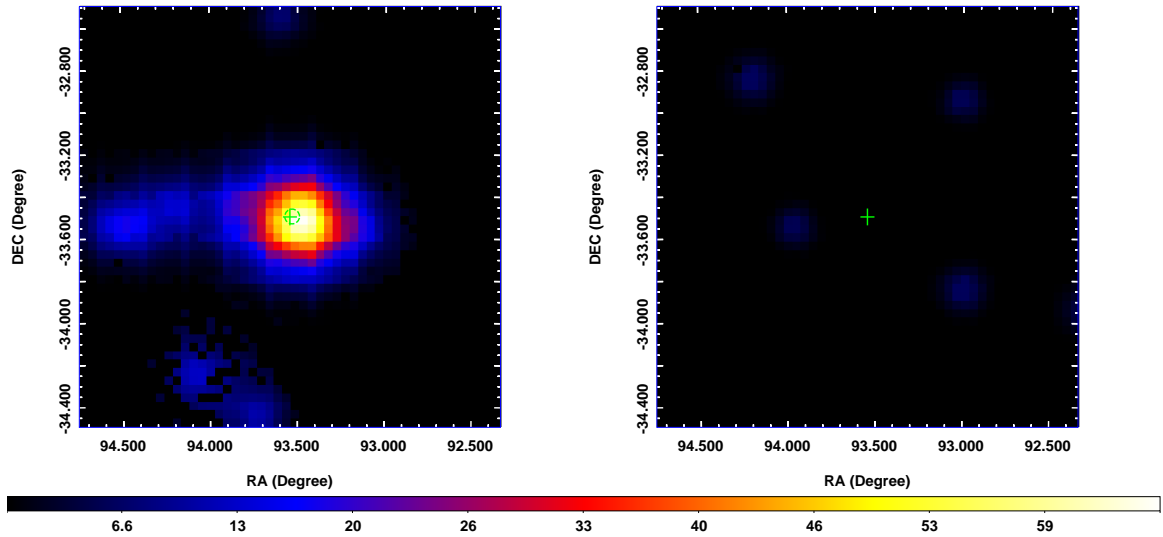}
\caption{TS maps of a $\mathrm{2^{o}\times2^{o}}$ region
centered at \psr\ in the 25--500 GeV band, during the onpulse ({\it left} panel)
 and offpulse ({\it right} panel) phase intervals. The image scale of the maps 
is 0\fdg04 pixel$^{-1}$. The color bar indicates the TS value range. The green 
crosses mark the position of \psr. The green circle marks 
the 2$\sigma$ error circle of the best-fit position obtained during 
the onpulse phase intervals.}
\label{fig:tsmap-phase}
\end{figure*}

We also performed likelihood analysis to the $>$25 GeV data during 
the onpulse and offpulse phase intervals, with the emission from the source 
modeled with a simple power law. The \gr\ emission from \psr\ was 
detected with TS$\simeq 65$, having $\Gamma= 2.8\pm$0.9 and 
photon flux 
$F_{25-500}= 1.0\pm 0.4\times 10^{-10}$ photons~s$^{-1}$\,cm$^{-2}$ 
during the onpulse phase intervals. During the offpulse phase intervals, 
the \gr\ emission was not detected (TS$\simeq 0$), and the derived 95\% photon 
flux upper limit is
$8\times 10^{-12}$ photons~s$^{-1}$\,cm$^{-2}$. 
Two TS maps during these two phase intervals are shown in 
Figure~\ref{fig:tsmap-phase}. 
We ran \textit{gtfindsrc} in the LAT software package to determine the 
position during the onpulse phase intervals and obtained R.A.=93\fdg53, 
Decl.= $-$33\fdg50, (equinox J2000.0), with 1$\sigma$ 
nominal uncertainty of 0\fdg02. \psr\ is 0\fdg01 from the best-fit position 
and within the 1$\sigma$ error circle. The result confirmed the detection 
of pulsed \gr\ emission from photon folding.

\section{Discussion}
\label{sec:disc}

\subsection{Spectral shape of MSPs}

Having analyzed approximately 7.5 yrs of \fermi\ LAT data for 39 MSPs
reported in 2PC, we have obtained their spectra, which are all well
described by a power law with exponential cutoff. We have thus determined their
general spectral shape by fitting the spectra with such a function. Due to
the intrinsic differences in their spectra, the allowed spectral shape region
for MSPs is relatively large. However this spectral shape can be used
to find candidate MSPs among the unidentified LAT sources. For example,
using the criteria of $>$5 degrees, significant curvature in a spectrum,
and non-variable, we \citep{dai+16} have found 24 such sources from
the \fermi\ LAT third source catalog \citep{3fgl15}, but two of them,
J0318.1+0252 and J2053.9+2922, likely have spectra not consistent with
the spectral shape of MSPs because either $\Gamma$ or $E_c$ found
for \gr\ emission of the two sources are not in the spectral shape range. 
Based on the known properties of the different
types of LAT sources, they could be MSPs with quite different
spectra or even other types of sources, for example the dark matter subhalo 
candidates as suggested by \citet{bhl15}. For
the purpose of finding candidate MSPs and other types of unidentified
\gr\ sources,  searching
through LAT sources at high Galactic latitudes by comparing their spectra 
with the spectral shape of MSPs we have determined is warranted.

\subsection{Pulsed \gr\ emission above 25 GeV from \psr}

High energy \gr\ emission is seen from 27 pulsars, as reported in
the first \fermi\ catalog of sources above 10 GeV \citep{ack10gev+13}.
Among them 20 sources 
were found to have pulsed \gr\ emission in $>$10 GeV band, including 17 young 
pulsars and 3 MSPs. Furthermore, \psr\ was one of 12 pulsars found to have \gr\ 
pulsations in $>$25 GeV band, although it was only marginally detected
in \citet{ack10gev+13}. Our analysis, likely due to the longer time
period of data (7.5 yrs vs. 3 yrs) and overall sensitivity improvement in 
the Pass 8 data, has
shown that there is significant pulsed \gr\ emission upto 200 GeV from
this MSP. The result has added \psr\ to the group of the Crab and Vela
pulsars that have been found to have $>$50 GeV pulsed
emission (e.g., \citealt{hk15} and references therein).

The mechanism of the very high-energy emission from pulsars remains 
to be solved.
Currently the inverse-Compton scattering 
process in the outer magnetosphere or the pulsar wind region 
is considered to produce the pulsed emission detected in $>$10 GeV band 
from the Crab pulsar (see, e.g., \citealt{ale+11,abk12,lyu13,hk15}). 
Alternatively a non-stationary outer gap scenario has also been
proposed recently \citep{tnc16}, which has been used to interpret 
the $>$50 GeV pulsed emission from the Vela pulsar \citep{leu+14}. 
In this scenario, the observed spectrum of a pulsar is the superposition 
of emission from the variable outer-gap structures. 
In Figure~\ref{fig:spectras}, we show the model fits 
to \gr\ emission from the Crab \citep{2fpsr13,ale+11} and 
Vela \citep{leu+14} pulsars 
(scaled by their 0.1--100 GeV total LAT fluxes respectively) 
for comparison. 
\psr\ possibly has a stronger $\sim$200 GeV component
than the Crab and Vela pulsars, although the large uncertainty does not
allow a clear conclusion to be drawn. In order to investigate the emission
process responsible for the high-energy component from \psr, detailed modeling
(such as those in \citealt{hk15,tnc16}) is needed.

\acknowledgements
This research made use of the High Performance Computing Resource in the Core
Facility for Advanced Research Computing at Shanghai Astronomical Observatory.
This research was supported by the Shanghai Natural Science 
Foundation for Youth (13ZR1464400), the National Natural Science Foundation
of China for Youth (11403075), the National Natural Science Foundation
of China (11373055), and the Strategic Priority Research Program
``The Emergence of Cosmological Structures" of the Chinese Academy
of Sciences (Grant No. XDB09000000). Z.W. acknowledges the support by 
the CAS/SAFEA International Partnership Program for Creative Research Teams.

\begin{table}
\tabletypesize{\footnotesize}
\tablewidth{240pt}
\caption{Likelihood analysis results for 39 MSP targets}
\label{tab:likelihood}
\begin{tabular}{lcccc}
\hline
Source name & $\Gamma$ & E$_{c}$ & F$_{100}$ & TS \\
 &  & (GeV) & (10$^{-12}$ erg cm$^{-2}$ s$^{-1}$) &  \\ 
\hline
J0023$+$0923 & 1.3$\pm$0.2 & 1.9$\pm$0.5 & 7$\pm$1 & 447 \\ 
J0030$+$0451 & 1.29$\pm$0.04 & 2.0$\pm$0.1 & 60$\pm$2 & 14973 \\ 
J0034$-$0534 & 1.58$\pm$0.08 & 3.2$\pm$0.5 & 18$\pm$1 & 2117 \\
J0101$-$6422 & 1.3$\pm$0.1 & 2.2$\pm$0.3 & 13$\pm$1 & 1835 \\
J0102$+$4839 & 1.73$\pm$0.09 & 6$\pm$1 & 16$\pm$1 & 1259 \\
J0218$+$4232 & 1.98$\pm$0.04 & 4.7$\pm$0.6 & 48$\pm$2 & 6417 \\
J0340$+$4130 & 1.16$\pm$0.09 & 3.4$\pm$0.4 & 20$\pm$2 & 2274 \\
J0437$-$4715 & 1.2$\pm$0.1 & 0.9$\pm$0.1 & 17$\pm$1 & 2830 \\
J0610$-$2100 & 1.4$\pm$0.2 & 2.1$\pm$0.5 & 8$\pm$1 & 477 \\
J0613$-$0200 & 1.42$\pm$0.07 & 2.9$\pm$0.3 & 31$\pm$2 & 2907 \\
J0614$-$3329 & 1.38$\pm$0.02 & 4.9$\pm$0.2 & 112$\pm$2 & 33600 \\
J0751$+$1807 & 1.4$\pm$0.1 & 3.6$\pm$0.7 & 13$\pm$1 & 1450 \\
J1024$-$0719 & 1.2$\pm$0.3 & 2.2$\pm$0.7 & 4$\pm$2 & 213 \\
J1124$-$3653 & 1.5$\pm$0.1 & 3.7$\pm$0.7 & 13$\pm$1 & 1088 \\
J1125$-$5825 & 1.7$\pm$0.2 & 7$\pm$3 & 9$\pm$1 & 230 \\
J1231$-$1411 & 1.09$\pm$0.03 & 2.4$\pm$0.1 & 100$\pm$2 & 28753 \\
J1446$-$4701 & 0.7$\pm$0.4 & 1.7$\pm$0.6 & 5$\pm$1 & 249 \\
J1514$-$4946 & 1.32$\pm$0.07 & 4.5$\pm$0.4 & 40$\pm$2 & 3831 \\
J1600$-$3053 & 0.5$\pm$0.2 & 2.5$\pm$0.5 & 7$\pm$2 & 487 \\
J1614$-$2230 & 0.8$\pm$0.1 & 2.0$\pm$0.2 & 25$\pm$2 & 2919 \\
J1658$-$5324$^{*}$ & 1.6$\pm$0.2 & 1.8$\pm$0.3 & 19$\pm$3 & 789 \\
J1713$+$0747 & 1.5$\pm$0.2 & 3.2$\pm$0.8 & 10$\pm$1 & 578 \\
J1741$+$1351 & 0.5$\pm$0.6 & 1.5$\pm$0.6 & 3$\pm$1 & 152 \\
J1744$-$1134 & 1.53$\pm$0.07 & 1.8$\pm$0.1 & 39$\pm$2 & 2439 \\
J1747$-$4036 & 1.2$\pm$0.2 & 2.3$\pm$0.5 & 9$\pm$1 & 274 \\
J1810$+$1744 & 2.08$\pm$0.07 & 5$\pm$1 & 24$\pm$1 & 1853 \\
J1823$-$3021A & 1.1$\pm$0.2 & 3.3$\pm$0.6 & 10$\pm$1 & 496 \\
J1858$-$2216$^{*}$ & 0.4$\pm$0.3 & 1.4$\pm$0.2 & 9$\pm$2 & 626 \\
J1902$-$5105 & 1.73$\pm$0.07 & 2.9$\pm$0.4 & 23$\pm$1 & 2502 \\
J1959$+$2048 & 1.6$\pm$0.1 & 2.5$\pm$0.5 & 14$\pm$1 & 532 \\
J2017$+$0603 & 1.07$\pm$0.08 & 3.8$\pm$0.4 & 34$\pm$2 & 4745 \\
J2043$+$1711 & 1.57$\pm$0.06 & 4.9$\pm$0.6 & 29$\pm$1 & 3437 \\
J2047$+$1053 & 1.3$\pm$0.4 & 3$\pm$1 & 4$\pm$2 & 152 \\
J2051$-$0827 & 0.8$\pm$0.4 & 2.0$\pm$0.7 & 3$\pm$1 & 125 \\
J2124$-$3358 & 0.85$\pm$0.07 & 1.8$\pm$0.1 & 38$\pm$2 & 7767 \\
J2214$+$3000 & 1.13$\pm$0.06 & 2.1$\pm$0.2 & 32$\pm$2 & 6015 \\
J2215$+$5135 & 1.2$\pm$0.1 & 3.6$\pm$0.7 & 12$\pm$1 & 817 \\
J2241$-$5236 & 1.36$\pm$0.05 & 3.0$\pm$0.3 & 32$\pm$2 & 7167 \\
J2302$+$4442 & 1.13$\pm$0.06 & 3.0$\pm$0.2 & 37$\pm$2 & 6157 \\
\hline
\end{tabular}
\vskip 1mm
\footnotesize{The results for the sources marked with \textquotedblleft\textasteriskcentered\textquotedblright\ were obtained in $>$0.2 GeV band.}
\end{table}

\begin{sidewaystable}
\tabletypesize{\footnotesize}
\tablewidth{240pt}
\caption{Spectral flux points for the MSP targets}
\label{tab:point}
\small
\begin{tabular}{lccccccccccc}
\hline
 & 0.13 & 0.22 & 0.38 & 0.65 & 1.10 & 1.88 & 3.21 & 5.48 & 9.34 & 15.93 & 27.16 \\
 & (GeV) & (GeV) & (GeV) & (GeV) & (GeV) & (GeV) & (GeV) & (GeV) & (GeV) & (GeV) & (GeV) \\ 
\hline
J0023$+$0923 & 2$\pm$1 & -- & 2.2$\pm$0.5 & 2.2$\pm$0.3 & 2.3$\pm$0.3 & 2.0$\pm$0.3 & 0.5$\pm$0.2 & -- & -- & -- & -- \\ 
J0030$+$0451 & 5$\pm$1 & 9.4$\pm$0.7 & 13.5$\pm$0.6 & 17.9$\pm$0.6 & 19.9$\pm$0.6 & 19.1$\pm$0.7 & 14.5$\pm$0.8 & 8.6$\pm$0.8 & 2.8$\pm$0.6 & 0.8$\pm$0.4 & -- \\
J0034$-$0534 & 2.2$\pm$0.9 & 3.4$\pm$0.5 & 4.4$\pm$0.4 & 4.8$\pm$0.4 & 4.8$\pm$0.4 & 5.2$\pm$0.4 & 4.1$\pm$0.4 & 2.3$\pm$0.4 & 1.5$\pm$0.4 & -- & -- \\
J0101$-$6422 & -- & 2.5$\pm$0.4 & 2.9$\pm$0.3 & 3.7$\pm$0.3 & 4.5$\pm$0.3 & 4.3$\pm$0.4 & 3.2$\pm$0.4 & 1.4$\pm$0.3 & 0.8$\pm$0.3 & -- & -- \\
J0102$+$4839 & 4$\pm$1 & 4$\pm$1 & 3.5$\pm$0.6 & 3.3$\pm$0.4 & 3.9$\pm$0.4 & 4.6$\pm$0.4 & 3.5$\pm$0.4 & 2.8$\pm$0.5 & 1.6$\pm$0.4 & 0.9$\pm$0.4 & -- \\
J0218$+$4232 & 13$\pm$2 & 13$\pm$1 & 13.8$\pm$0.7 & 12.4$\pm$0.6 & 12.1$\pm$0.5 & 9.3$\pm$0.5 & 6.9$\pm$0.5 & 5.0$\pm$0.6 & 2.1$\pm$0.5 & 1.4$\pm$0.5 & -- \\
J0340$+$4130 & -- & -- & 3.0$\pm$0.5 & 3.9$\pm$0.4 & 5.6$\pm$0.4 & 6.3$\pm$0.4 & 7.3$\pm$0.5 & 4.8$\pm$0.6 & 3.7$\pm$0.6 & 0.7$\pm$0.4 & -- \\
J0437$-$4715 & 4.8$\pm$0.8 & 5.1$\pm$0.5 & 5.2$\pm$0.4 & 6.4$\pm$0.3 & 6.1$\pm$0.3 & 3.7$\pm$0.3 & 1.9$\pm$0.3 & -- & -- & -- & -- \\
J0610$-$2100 & -- & 3.0$\pm$0.7 & 1.5$\pm$0.4 & 2.5$\pm$0.3 & 2.5$\pm$0.3 & 2.3$\pm$0.3 & 1.3$\pm$0.3 & 1.3$\pm$0.3 & -- & -- & -- \\
J0613$-$0200 & -- & 5$\pm$1 & 7.8$\pm$0.9 & 9.7$\pm$0.8 & 9.3$\pm$0.9 & 10.2$\pm$0.7 & 8.3$\pm$0.7 & 6.2$\pm$0.7 & 2.4$\pm$0.6 & -- & -- \\
J0614$-$3329 & 8$\pm$1 & 11.7$\pm$0.6 & 16.0$\pm$0.5 & 21.7$\pm$0.6 & 27.9$\pm$0.7 & 32.9$\pm$0.9 & 34$\pm$1 & 28$\pm$1 & 19$\pm$1 & 9$\pm$1 & 4$\pm$1 \\
J0751$+$1807 & 4.9$\pm$0.8 & 3.3$\pm$0.6 & 1.8$\pm$0.4 & 2.5$\pm$0.3 & 3.4$\pm$0.3 & 4.7$\pm$0.4 & 3.5$\pm$0.4 & 3.3$\pm$0.5 & 0.9$\pm$0.3 & -- & -- \\
J1024$-$0719 & -- & -- & 1.0$\pm$0.3 & -- & 1.5$\pm$0.2 & 1.3$\pm$0.2 & 1.3$\pm$0.3 & 0.7$\pm$0.3 & -- & -- & -- \\
J1124$-$3653 & 3$\pm$1 & 2.4$\pm$0.9 & 2.7$\pm$0.5 & 2.8$\pm$0.4 & 3.4$\pm$0.3 & 3.7$\pm$0.4 & 3.8$\pm$0.4 & 2.6$\pm$0.4 & 0.8$\pm$0.3 & -- & -- \\
J1125$-$5825 & -- & -- & -- & 3.4$\pm$0.7 & 3.0$\pm$0.5 & 1.6$\pm$0.4 & 2.7$\pm$0.5 & 2.1$\pm$0.5 & -- & 0.9$\pm$0.5 & -- \\
J1231$-$1411 & 3$\pm$2 & 9.8$\pm$0.8 & 15.7$\pm$0.6 & 25.1$\pm$0.6 & 32.0$\pm$0.8 & 34.8$\pm$0.9 & 32$\pm$1 & 22$\pm$1 & 9$\pm$1 & 2.7$\pm$0.7 & -- \\
J1446$-$4701 & -- & -- & -- & 1.3$\pm$0.4 & 1.9$\pm$0.3 & 2.5$\pm$0.3 & 2.1$\pm$0.4 & 0.8$\pm$0.3 & 0.7$\pm$0.3 & -- & -- \\
J1514$-$4946 & -- & 6$\pm$2 & 5.4$\pm$0.9 & 8.1$\pm$0.6 & 10.7$\pm$0.6 & 10.9$\pm$0.6 & 12.8$\pm$0.8 & 10.6$\pm$0.9 & 8.2$\pm$0.9 & 2.5$\pm$0.7 & -- \\
J1600$-$3053 & -- & -- & -- & -- & 2.1$\pm$0.3 & 2.6$\pm$0.3 & 2.9$\pm$0.4 & 2.4$\pm$0.4 & 1.7$\pm$0.4 & -- & -- \\
J1614$-$2230 & -- & -- & 3.7$\pm$0.8 & 5.7$\pm$0.5 & 7.5$\pm$0.5 & 10.6$\pm$0.6 & 9.4$\pm$0.6 & 5.4$\pm$0.6 & 1.9$\pm$0.5 & 1.7$\pm$0.6 & -- \\
J1658$-$5324 & -- & 2.9$\pm$0.4 & 5.8$\pm$0.9 & 6.0$\pm$0.6 & 6.3$\pm$0.5 & 4.5$\pm$0.5 & 2.8$\pm$0.4 & 1.5$\pm$0.4 & -- & -- & -- \\
J1713$+$0747 & -- & 2.4$\pm$0.7 & 3.1$\pm$0.5 & 2.4$\pm$0.4 & 2.5$\pm$0.3 & 3.1$\pm$0.3 & 3.2$\pm$0.4 & 1.4$\pm$0.4 & 0.8$\pm$0.3 & -- & -- \\
J1741$+$1351 & -- & -- & -- & 0.9$\pm$0.3 & 1.3$\pm$0.3 & 1.2$\pm$0.3 & 1.7$\pm$0.3 & 0.6$\pm$0.2 & -- & -- & -- \\
J1744$-$1134 & 7$\pm$4 & 9$\pm$1 & 10.1$\pm$0.9 & 13.2$\pm$0.9 & 14.4$\pm$0.9 & 10.4$\pm$0.6 & 7.1$\pm$0.6 & 2.1$\pm$0.5 & -- & -- & -- \\
J1747$-$4036 & -- & 3$\pm$2 & 3.5$\pm$0.8 & 3.0$\pm$0.6 & 3.2$\pm$0.5 & 2.4$\pm$0.4 & 2.7$\pm$0.4 & 1.5$\pm$0.4 & -- & -- & -- \\
J1810$+$1744 & 4$\pm$2 & 6.2$\pm$0.8 & 7.0$\pm$0.5 & 7.7$\pm$0.4 & 5.9$\pm$0.4 & 4.6$\pm$0.4 & 3.3$\pm$0.4 & 1.5$\pm$0.3 & 1.2$\pm$0.4 & -- & -- \\
J1823$-$3021A & -- & 3$\pm$1 & 2.5$\pm$0.7 & 1.8$\pm$0.5 & 3.5$\pm$0.4 & 3.3$\pm$0.4 & 3.3$\pm$0.4 & 3.3$\pm$0.5 & 1.0$\pm$0.4 & -- & -- \\
J1858$-$2216 & -- & -- & -- & 2.7$\pm$0.4 & 3.4$\pm$0.4 & 4.3$\pm$0.4 & 3.8$\pm$0.4 & 1.5$\pm$0.4 & 0.9$\pm$0.3 & -- & -- \\
J1902$-$5105 & 3.6$\pm$0.8 & 5.7$\pm$0.5 & 6.6$\pm$0.4 & 7.5$\pm$0.4 & 6.6$\pm$0.4 & 4.8$\pm$0.4 & 3.8$\pm$0.4 & 2.4$\pm$0.4 & 1.1$\pm$0.4 & -- & -- \\
J1959$+$2048 & -- & -- & 3$\pm$1 & 5.2$\pm$0.7 & 4.8$\pm$0.6 & 4.3$\pm$0.5 & 2.9$\pm$0.4 & 1.2$\pm$0.3 & -- & -- & -- \\
J2017$+$0603 & -- & 3.6$\pm$0.6 & 6.1$\pm$0.5 & 8.9$\pm$0.5 & 11.2$\pm$0.6 & 11.4$\pm$0.7 & 11.1$\pm$0.8 & 7.6$\pm$0.9 & 2.8$\pm$0.7 & -- & -- \\
J2043$+$1711 & 3$\pm$1 & 5.1$\pm$0.7 & 5.5$\pm$0.5 & 6.3$\pm$0.4 & 8.2$\pm$0.5 & 7.3$\pm$0.5 & 7.9$\pm$0.6 & 5.9$\pm$0.6 & 3.6$\pm$0.6 & 1.1$\pm$0.5 & -- \\
J2047$+$1053 & -- & -- & -- & -- & 0.7$\pm$0.2 & 1.3$\pm$0.3 & 1.3$\pm$0.3 & 0.9$\pm$0.3 & -- & -- & -- \\
J2051$-$0827 & -- & -- & -- & -- & 1.3$\pm$0.3 & 0.9$\pm$0.2 & 0.8$\pm$0.2 & 0.9$\pm$0.3 & -- & -- & -- \\
J2124$-$3358 & -- & 3.4$\pm$0.7 & 6.3$\pm$0.5 & 9.6$\pm$0.5 & 12.8$\pm$0.5 & 15.0$\pm$0.6 & 13.1$\pm$0.7 & 8.0$\pm$0.7 & 2.3$\pm$0.5 & -- & -- \\
J2214$+$3000 & -- & 4.8$\pm$0.6 & 6.1$\pm$0.5 & 9.0$\pm$0.4 & 10.7$\pm$0.5 & 11.4$\pm$0.5 & 8.8$\pm$0.6 & 6.7$\pm$0.6 & 1.3$\pm$0.4 & -- & -- \\
J2215$+$5135 & -- & -- & 1.5$\pm$0.5 & 2.7$\pm$0.4 & 4.5$\pm$0.4 & 3.5$\pm$0.4 & 3.5$\pm$0.4 & 2.9$\pm$0.5 & 2.3$\pm$0.5 & 0.9$\pm$0.4 & -- \\
J2241$-$5236 & 3.4$\pm$0.9 & 5.3$\pm$0.5 & 6.4$\pm$0.5 & 7.8$\pm$0.4 & 10.5$\pm$0.4 & 9.7$\pm$0.5 & 9.8$\pm$0.6 & 5.9$\pm$0.6 & 3.0$\pm$0.6 & -- & 1.3$\pm$0.7 \\
J2302$+$4442 & -- & 2.9$\pm$0.6 & 5.4$\pm$0.5 & 7.9$\pm$0.4 & 10.6$\pm$0.5 & 12.3$\pm$0.6 & 12.7$\pm$0.7 & 9.8$\pm$0.8 & 4.6$\pm$0.7 & 1.7$\pm$0.6 & -- \\
\hline
\end{tabular}
\vskip 1mm
\footnotesize{Fluxes are in units of 10$^{-12}$ erg cm$^{-2}$ s$^{-1}$.}
\end{sidewaystable}

\begin{table}
\tabletypesize{\footnotesize}
\tablewidth{240pt}
\setlength{\tabcolsep}{2pt}
\caption{Exponentially cutoff power-law fits for \psr.}
\label{tab:likelihoods}
\centering
\begin{tabular}{lccccc}
\hline
Data set & $>$0.1 GeV Flux & $\Gamma$ & E$_{c}$ & b & TS \\
 & (10$^{-8}$ photon cm$^{-2}$ s$^{-1}$) &  & (GeV) &  &  \\
\hline
Total data & 8.6 $\pm$ 0.2 & 1.38 $\pm$ 0.02 & 4.9 $\pm$ 0.2 & 1 & 33576 \\
           & 8.2 $\pm$ 0.2 & 1.1 $\pm$ 0.1 & 1.6 $\pm$ 0.9 & 0.64 $\pm$ 0.09 & 33566 \\
\hline
Onpulse data & 19.1 $\pm$ 0.3 & 1.29 $\pm$ 0.02 & 5.1 $\pm$ 0.3 & 1 & 37013 \\
             & 18.5 $\pm$ 0.4 & 1.1 $\pm$ 0.1 & 2.0 $\pm$ 0.9 & 0.68 $\pm$ 0.09 
& 36997 \\
\hline
Offpulse data & 3.7 $\pm$ 0.2 & 1.42 $\pm$ 0.07 & 2.1 $\pm$ 0.2 & 1 & 3766 \\
\hline
\end{tabular}
\vskip 1mm
\end{table}

\clearpage
\begin{table}
\tabletypesize{\footnotesize}
\tablecolumns{10}
\tablewidth{240pt}
\setlength{\tabcolsep}{2pt}
\caption{\fermi\ LAT flux measurements of \psr}
\label{tab:spectra}
\centering
\begin{tabular}{lccccccc}
\hline
\multicolumn{2}{c}{ } &
\multicolumn{2}{c}{Total} &
\multicolumn{2}{c}{Onpulse} &
\multicolumn{2}{c}{Offpulse}  \\
\hline
$E$ & Band & $E^2dN(E)/dE$ & TS & $E^2dN(E)/dE$ & TS & $E^2dN(E)/dE$ & TS \\
(GeV) & (GeV) & (10$^{-12}$ erg cm$^{-2}$ s$^{-1}$) &  & (10$^{-12}$ erg cm$^{-2}$ s$^{-1}$) &  & (10$^{-12}$ erg cm$^{-2}$ s$^{-1}$) &  \\ \hline
0.14 & 0.1--0.2 & 8.5$\pm$0.7 & 276 & 17$\pm$1 & 310 & 4.8$\pm$0.8 & 63 \\
0.29 & 0.2--0.4 & 13.8$\pm$0.5 & 1558 & 30$\pm$1 & 1868 & 6.4$\pm$0.5 & 265 \\
0.59 & 0.4--0.8 & 20.8$\pm$0.5 & 5119 & 47$\pm$1 & 5536 & 8.8$\pm$0.4 & 865 \\
1.20 & 0.8--1.7 & 28.8$\pm$0.6 & 9323 & 71$\pm$2 & 9877 & 9.6$\pm$0.5 & 1303 \\
2.44 & 1.7--3.5 & 33.5$\pm$0.8 & 9731 & 88$\pm$2 & 10298 & 8.5$\pm$0.5 & 1013 \\
4.96 & 3.5--7.1 & 30$\pm$1 & 5526 & 86$\pm$3 & 6301 & 4.0$\pm$0.5 & 259 \\
10.08 & 7.1--14.4 & 17$\pm$1 & 1678 & 53$\pm$4 & 2045 & 1.1$\pm$0.4 & 30 \\
20.50 & 14.4--29.2 & 7$\pm$1 & 333 & 23$\pm$3 & 439 & 0.3 & 0 \\
41.70 & 29.2--59.5 & 0.6$\pm$0.5 & 6 & 2$\pm$1 & 10 & 0.9 & 0 \\
84.79 & 59.5--120.9 & 0.6$\pm$0.6 & 6 & 2$\pm$2 & 8 & 1.2 & 0 \\
172.42 & 120.9--245.9 & 2$\pm$2 & 5 & 8$\pm$6 & 9 & 2.5 & 0 \\
350.62 & 245.9--500.0 & 3.8 & 0 & 14.3 & 0 & 5.1 & 0 \\
\hline
\end{tabular}
\vskip 1mm
\footnotesize{Note: fluxes with uncertainties are given in energy bins with 
$>$2$\sigma$ detection significance, and fluxes without uncertainties are 
the 95$\%$ upper limits.}
\end{table}

\end{document}